\documentclass[aps,pre,showpacs,preprint,groupedaddress]{revtex4-1}

\usepackage{graphicx}
\usepackage{amsmath}
\usepackage{hyperref}

\begin{document}


\title{Stability of paramagnetic spheroid in precessing field}


\author{J\={a}nis C\={\i}murs}
\author{Artis Brasovs}
\author{Kaspars \={E}rglis}
\altaffiliation{Faculty of Physics, Mathematics and Optometry, University of Latvia, Ze\c{l}\c{l}u 23, R\={\i}ga, LV-1001, Latvia}
\email[\\E-mail: ]{janis.cimurs@lu.lv}
\homepage[Homepage: ]{mmml.lu.lv}
\affiliation{Faculty of Physics, Mathematics and Optometry, University of Latvia}


\date{\today}

\begin{abstract}
Stability analysis of paramagnetic prolate and oblate spheroidal particle in precessing magnetic field is studied. Bifurcation diagram is calculated analytically in dependence of magnetic field frequency and precession angle. The direction of particle in synchronous regime is calculated. Rotational dynamics and mean rotational frequency in asynchronous regime is found. Obtained theoretical model gives possibility to calculate analytically dynamics of the particle in limiting case (when motion is periodic). Theoretically obtained models were compared with experimental results of rod like particle dynamics in precessing magnetic field. Experimental results are in good agreement with proposed theory.
\end{abstract}


\maketitle

\section{Introduction}
\label{intro}
Recently active topic of research is magnetic particle structure formation in different field configurations \cite{AdvancesSuper,KLAPP2016}, which has been investigated experimentally, numerically and theoretically. Great interest in magnetic particles is due to possible applications. For example paramagnetic particles can be used to measure rheological properties of the fluid \cite{SoftMatter}. Using external magnetic field we can tune interaction between paramagnetic particles as can be seen in \cite{TiernoInteract}. By tuning interaction between paramagnetic particles by changing properties of precessing magnetic field we can form chains \cite{MartinChain} or if we use more complex particles like Janus particles we can obtain layers or tubes \cite{JanusNature}. Magnetic Janus particles posses magnetic anisotropy and therefore can be modeled as oblate paramagnetic particles.

To fully understand these structure formation of magnetic particles, we need to start by fully understanding dynamics of individual particle. Previous studies have found experimentally the transition between synchronous regime and asynchronous regime for prolate paramagnetic particle in precessing field \cite{CebersElipse}. Numerical studies approved experimentally obtained results and expanded bifurcation diagram including oblate paramagnetic particle and wider range of field parameters \cite{superCimurs}. Nevertheless full dynamics of the particle in asynchronous regime was not investigated. 

In this work we formulate a model of the dynamics of the spheroidal paramagnetic particle which is applicable both for prolate and oblate particles, analytically obtain bifurcation diagram and calculate trajectory of the particle in stable \lq\lq{}periodic\rq\rq{} motion. Due to symmetry, all obtained equations can be used both for prolate and oblate particle. The only difference is the region of applicability, which is defined by bifurcation diagram. Although the equations for prolate and oblate particles are the same, due to different regions of applicability, the observable behavior differs.

\section{Theoretical model}
 
 A spheroidal paramagnetic particle has anisotropic magnetic susceptibility due to the anisotropy of its demagnetizing field factors $N_{\|,\perp}$ \cite{OsbornDemag,StonerDemag}. If we denote the symmetry axis of the particle by $\vec n$, then magnetic moment $\vec m$ in an external magnetic field $\vec H=H\vec h$ reads
\begin{equation}
  \vec m=VH\left[\chi_\perp\vec h+\left(\chi_\|-\chi_\perp\right)\left(\vec n\cdot\vec h\right)\vec n\right]\text{ ,}
\end{equation}
where $V$ is volume of the particle and $\chi_{\|,\perp}$ are magnetic susceptibilities, which depend on the demagnetizing field factors according to \cite{Landau} $\chi_{\|,\perp}=\chi_0/(1+\chi_0N_{\|,\perp})$, where $\chi_0$ is the isotropic intrinsic susceptibility of the particle. The particle is in the fluid which rotates with angular velocity $\vec\omega_H=\omega_H\vec e_H$.

\begin{figure}[ht]
                \includegraphics[width=0.25\textwidth]{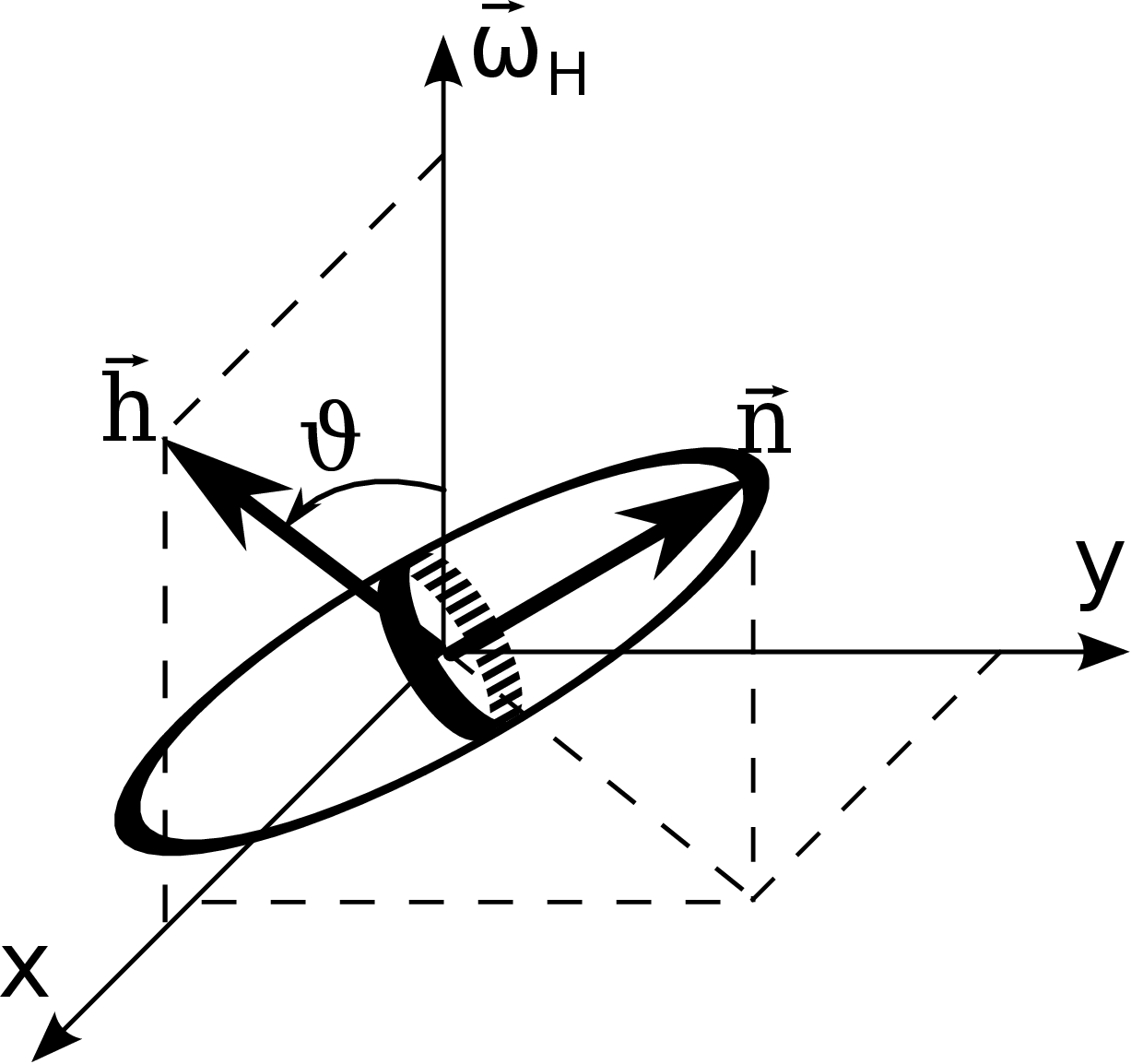}
                \caption{Schematic picture of the problem. Particle with direction $\vec n$ in external magnetic field with direction $\vec h$ and rotating fluid with angular velocity $\vec \omega_H$}
                \label{fig:prez}
\end{figure}

If we are interested in rotational dynamics of magnetic particle in precessing magnetic field which precesses with angular velocity $\vec\omega_H$ in stationary fluid, then this is equivalent to look at particle in stationary magnetic field and rotating fluid with angular velocity $-\vec\omega_H$.

The balance of the viscous and magnetic torque (for low Reynolds number) reads:
\begin{equation}
  V\left(\chi_\|-\chi_\perp\right)H^2\left(\vec n\cdot\vec h\right)\vec n\times\vec h=\xi_r\left(\vec n\times\dot{\vec n}+\omega_H\vec e_H\right)\text{ ,}
\end{equation}
where $\xi_r$ is a rotational drag coefficient. $\xi_r$ for spheroidal particles can be found in \cite{CebersElipse}.

If we introduce anisotropy frequency $\omega_a=V H^2(\chi_\|-\chi_\perp)/\xi_r$, then equation of the rotation of the particle can be written in the form:
\begin{equation}\label{eq:dndt}
  \dfrac{d\vec n}{dt}=\omega_a\left(\vec n\cdot\vec h\right)\left[\vec h-\left(\vec n\cdot\vec h\right)\vec n\right]+\omega_H\vec e_H\times\vec n
\end{equation}
In general the angle between $\vec h$ and $\vec e_H$ is arbitrary. For prolate particle $\omega_a>0$, and for oblate $\omega_a<0$.

\subsection{Fixed points and their stability}

Setting $\dot{\vec n}=0$ let us find fixed points. By simple calculations, which can be found in appendix \ref{apx:cubic}, cubic equation can be found, whose roots determine fixed points:
\begin{equation}\label{eq:cubic}
  f(y)=y^3-y^2+\omega^2y-\omega^2\sigma^2\text{ ,}
\end{equation}
where $y=(\vec n\cdot\vec h)^2$, $\omega=\dfrac{\omega_H}{\omega_a}$ and $\sigma=\vec h\cdot\vec e_H$. The last term of the cubic equation \eqref{eq:cubic} is negative, that means that product of all roots is positive, therefore at least one root is positive. Since $y=(\vec n\cdot\vec h)^2$, this gives that for positive $y$ there exists fixed points. We can conclude that for every value of $\omega$ and $\sigma$ there exists fixed point.

Equation \eqref{eq:cubic} does not change if the sign of the $\omega$ (or $\omega_a$) is changed. This means that fixed points for prolate ($\omega_a>0$) and oblate ($\omega_a<0$) particles are the same. The only difference is stability of these points, which depend on the sign of $\omega_a$.

Fixed point in situation, where magnetic field is kept constant and magnetic fluid rotates corresponds to synchronous rotation of the particle with the field frequency if fluid is stationary and magnetic field precesses (rotates).

In order to investigate stability of fixed point (synchronous regime) we introduce small perturbation $\vec \varepsilon$ from the synchronous state $\vec n$. Since $\vec n$ is unit vector and its length can not change $\vec \varepsilon$ should be perpendicular to $\vec n$ and can be written in form:
\begin{equation}\label{eq:varepsilon}
  \vec\varepsilon=\varepsilon_1[\vec n\times\vec h]+\varepsilon_2[\vec e_H\times\vec n]\text{ ,}
\end{equation}
where $\varepsilon_1$ and $\varepsilon_2$ are small perturbations in two non-collinear directions.

After some arithmetical manipulations shown in appendix \ref{apx:pert} we can find time dependence of $\vec\varepsilon$:
\begin{equation}\label{eq:de1de2}
\left\{\begin{aligned}
  \frac{d\varepsilon_1}{dt}&=-\omega_a (\vec n\cdot\vec h)^2\varepsilon_1-\omega_a(\vec e_H\cdot\vec n)(\vec n\cdot\vec h)\varepsilon_2\\
  \frac{d\varepsilon_2}{dt}&=\frac{\omega_H^2}{\omega_a}\frac{(\vec e_H\cdot\vec n)}{(\vec n\cdot\vec h)}\varepsilon_1+\omega_a\left(2-(\vec n\cdot\vec h)^2\right)\varepsilon_2
\end{aligned}\right.
\end{equation}

The stability of the regime is determined by eigenvalues of Jacobi matrix \cite{Strogatz} and gives:
\begin{equation}\label{eq:lambda}
  \lambda_{1,2}=\omega_a\dfrac{\tau\pm\sqrt{\tau^2-4\Delta}}{2}
\end{equation}
In this situation $\Delta$ and $\tau$ can be calculated from first and second derivative of the cubic equation \eqref{eq:cubic}:
\[
  \Delta=\omega^2-2y+3y^2=f'(y)\qquad\tau=1-3y=-\frac12f''(y)\text{ .}
\]
Therefore it is sufficient to look at the first and second derivative of cubic function $f(y)$ at its roots $f(y)=0$ to examine the stability of fixed points.
The possible $\tau$ and $\Delta$ values and consequences to the fixed point $f(y)=0$ are  following:
\begin{itemize}
  \item If $\tau<0$ ($f''(y)>0$) and $\Delta>0$ ($f'(y)>0$), then $\dfrac{\lambda_1}{\omega_a}<0$ and $\lambda_2/\omega_a<0$ \eqref{eq:lambda}, therefore prolate particle ($\omega_a>0$) is stable in this point, but oblate particle ($\omega_a<0$) is unstable.
  \item If $\tau>0$ ($f''(y)<0$) and $\Delta>0$ ($f'(y)>0$), then $\lambda_1/\omega_a>0$ and $\lambda_2/\omega_a>0$ \eqref{eq:lambda}, therefore prolate particle ($\omega_a>0$) is unstable in this point, but oblate particle ($\omega_a<0$) is stable.
  \item If $\Delta<0$ ($f'(y)<0$), then $\lambda_1/\omega_a>0$ and $\lambda_2/\omega_a<0$ \eqref{eq:lambda}, it is saddle point, therefore both prolate particle ($\omega_a>0$) and oblate particle ($\omega_a<0$) are unstable in this point.
\end{itemize}

The coefficient at $y^3$ term in $f(y)$ is positive, therefore $f(y)$ is mainly increasing and if $f(y)=0$ has only real one root then  $f'(y)>0$ at that root. The change in stability at that point is determined from $f''(y)$ which changes sign at $y=\frac{1}{3}$. Putting this in $f(y)=0$ we get neutral stability curve:
\begin{equation}\label{eq:h13}
  \omega^2=\left(\frac{\omega_H}{\omega_a}\right)^2=\frac{2}{9\left(1-3\sigma^2\right)}
\end{equation}
This equation has real solution only if $\sigma<\frac{1}{\sqrt{3}}$. Otherwise $f(y)$ has one root with $f''(y)>0$ what means that prolate particle ($\omega_a>0$) has stable fixed point and oblate particle ($\omega_a<0$) has unstable.

If the equation $f(y)=0$ has 3 real roots, then between these 3 real roots one must have $f'(y)>0$ and $f''(y)<0$, one must have $f'(y)>0$ and $f''(y)>0$ and one must have $f'(y)<0$, which are between first two. therefore both particles prolate $\omega_a>0$ and oblate $\omega_a<0$ have one stable fixed point if $f(y)=0$ has 3 real roots. The equation \eqref{eq:cubic} has 3 real solution if:
\[
  18\omega^4\sigma^2-4\omega^2\sigma^2+\omega^4-4\omega^6-27\omega^4\sigma^4>0
\]
 Therefore \eqref{eq:cubic} has three solutions if $\omega_1(\sigma)<\omega<\omega_2(\sigma)$, where
\begin{equation}\label{eq:omega12}
  \omega_{1,2}(\sigma)=\sqrt{\frac{1+18\sigma^2-27\sigma^4\pm\sqrt{(1-9\sigma^2)^3(1-\sigma^2)}}{8}}\text{ .}
\end{equation}
This equation has real solution if $0<\sigma<\frac{1}{3}$.

If we put $\sigma=\frac{1}{3}$ in equation \eqref{eq:omega12} and \eqref{eq:h13} we get the same value $\omega=\frac{1}{\sqrt{3}}$. In this situation $f(y)$ becomes $f(y)=\left(y-\frac{1}{3}\right)^3$.

\begin{figure}
  \includegraphics[width=0.48\textwidth]{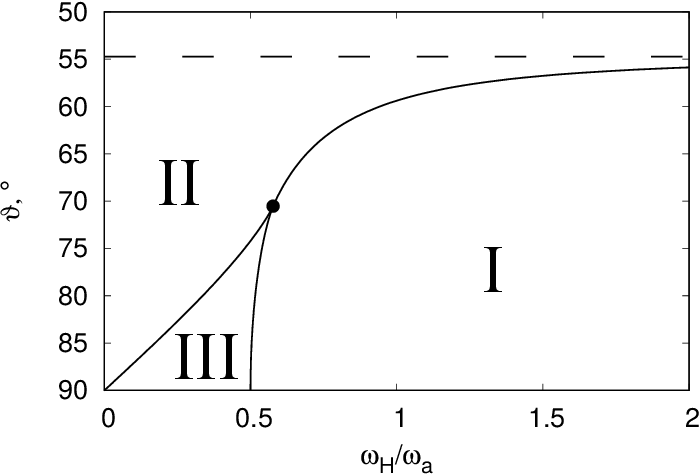}
\caption{\textbf{Bifurcation diagram} in dependence of fluid rotational frequency $\omega_H$ and angle $\vartheta$ between fluid rotation axis $\vec e_H$ and magnetic field axis $\vec h$. The prolate particle has a stable synchronous regime in region $II\bigcup III$, and the oblate particle in region $I\bigcup III$. The regions is divided by the neutral stability (solid) line. Long dashed line is asymptote at $\dfrac{\omega_H}{\omega_a}\rightarrow\infty$. The point shows a codimension-2 bifurcation point.}
\label{fig:Phase}
\end{figure}

The results of eq. \eqref{eq:h13} and \eqref{eq:omega12} can be visualized by bifurcation diagram fig. \ref{fig:Phase}, where $\vartheta$ is angle between $\vec h$ and $\vec e_H$ ($\vec h\cdot\vec e_H=\cos\vartheta$). In the bifurcation diagram region $I\bigcup II$ $f(y)$ has one root, where in region $I$ prolate particle has stable fixed point, but in region $II$ oblate particle has stable fixed point, but in region $III$ $f(y)$ has three roots therefore prolate and oblate particle have stable fixed points.

\subsection{Limit circle and rotational frequency}

In the situation, where prolate particle does not have stable fixed points, there should be a limit circle. Luckily in this situation limit circle is \lq\lq{}round\rq\rq{} and in this situation $\vec n$ rotates in plane perpendicular to some stationary vector $\vec e_L$. As shown in appendix \ref{apx:circle}, the equation for $\vec e_L$ is the same as for $\vec n$ \eqref{eq:dndt}, but with opposite sign at $\omega_a$. This gives that limit circle for prolate particle is stable if oblate particle has stable fixed point. Corresponding fixed point of oblate particle is fixed point of the vector $\vec e_L$ for prolate particle, which defines the plane of the rotation. The same reasoning is true for oblate particle without fixed points.

We also see that $\vec n$ of prolate particle is always perpendicular to $\vec e_L$, which is $\vec n$ of oblate particle. Therefore in situation where eq. \eqref{eq:cubic} has three roots (region $III$ in fig. \ref{fig:Phase}) $\vec n$ of prolate particle is perpendicular to $\vec n$ of oblate particle  (with equal $|\omega_a|$).

The rotation around limit circle is not uniform, but is periodic with mean angular velocity \eqref{eq:omega_L_vid}:
\begin{equation}\label{eq:omegaL}
  \bar \omega_L=\sqrt{\omega_H^2-\frac{\omega_a^2}{4}+\omega_a^2\frac{y}{2}\left(\frac{3y}{2}-1\right)}\text{ ,}
\end{equation}
where $y$ is the real root of \eqref{eq:cubic}.

It can be checked that mean angular velocity \eqref{eq:omegaL} is zero at boundary of region $III$ fig. \ref{fig:Phase}, but is not zero at boundary between regions $I$ and $II$ fig. \ref{fig:Phase}. For prolate particle if we slowly go from region $III$ into region $I$ than mean rotational frequency would continuously increase; same for oblate particle if it slowly goes from region $III$ into region $II$. If Prolate particle slowly goes from region $II$ into region $I$ than there would be jump in mean rotational frequency at the boundary between regions; same for oblate particle going from region $I$ into region $II$. The stable limit circle or fixed point near boundary between regions $I$ and $II$ is not very attractive, therefore it could take long time to reach stable state. But on the boundary of region $III$, the fixed point is already on the limit circle, therefore the stable state would be reached much faster.

\begin{figure}
  \includegraphics[width=0.48\textwidth]{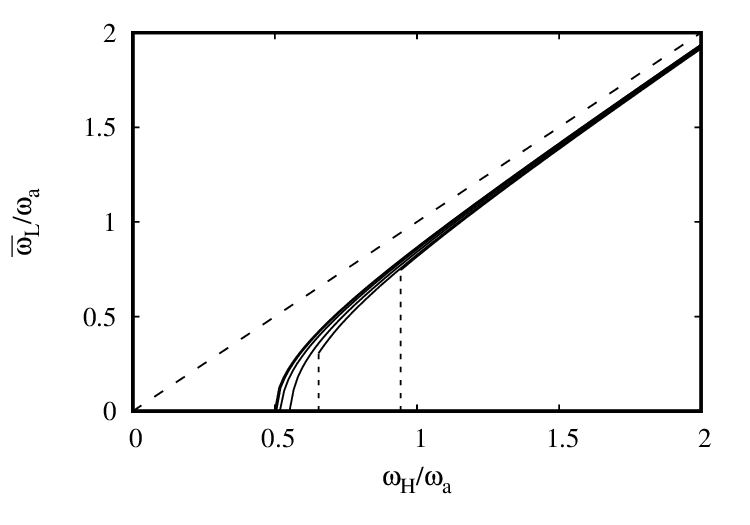}
\caption{Mean rotational frequency $\bar\omega_L$ of the prolate particle in dependence of rotational frequency of the fluid $\omega_H$ for different angle between magnetic field $\vec h$ and axis of the angular velocity $e_H$ for $\sigma=\vec h\cdot\vec e_H$ values 0, 0.1, 0.2, 0.3, 0.4 and 0.5. $\sigma$ value increases to the left. Long dashed line is rotational frequency of the fluid. Short dashed line shows jumps in the mean rotational frequency.}
\label{fig:omegaL}
\end{figure}

\begin{figure}
                \includegraphics[width=0.48\textwidth]{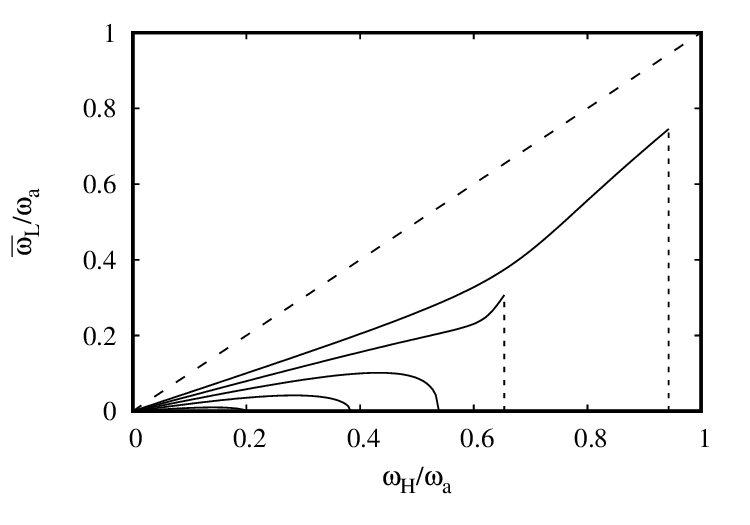}
                \caption{Mean rotational frequency $\bar\omega_L$ of the oblate particle in dependence of rotational frequency of the fluid $\omega_H$ for different angle between magnetic field $\vec h$ and axis of the angular velocity $e_H$ for $\sigma=\vec h\cdot\vec e_H$ values 0.1, 0.2, 0.3, 0.4 and 0.5. $\sigma$ value increases to the left. Long dashed line is rotational frequency of the fluid. Short dashed line shows jumps in the mean rotational frequency.}
                \label{fig:omegaL_ob}
\end{figure}

The mean angular velocity dependence of the fluid rotation is shown in fig. \ref{fig:omegaL} for prolate particle and for oblate particle it is shown in fig. \ref{fig:omegaL_ob}. The place of the jump in mean angular velocity is found from \eqref{eq:h13} and the height of the jump can be found by setting $y=\frac{1}{3}$, which gives:
\[
  \Delta\bar\omega_L=\sqrt{\omega_H^2-\dfrac{\omega_a^2}{3}}=\omega_H\sqrt{\dfrac{9}{2}\sigma^2-\dfrac{1}{2}}\text{ .}
\]

If instead of rotating fluid, we will put particle in stationary fluid, but magnetic field rotates, we get that particle rotates asynchronously in this situation. And particle lags behind magnetic field with average angular velocity $\omega_L$. Therefore average rotational frequency of spheroidal magnetic particle in precessing magnetic field in asynchronous regime is
\begin{equation}\label{eq:omegan}
  \bar \omega_n=\omega_H-\sqrt{\omega_H^2-\frac{\omega_a^2}{4}+\omega_a^2\frac{y}{2}\left(\frac{3y}{2}-1\right)}
\end{equation}

\section{Experiment}

To test provided theory, the experiment was made. In experiment superparamagnetic rods, which were synthesized according to the method detailed in \cite{rod}, was immersed in stationary fluid and subjected to precessing magnetic field. Precessing field consists of constant field with strength $H_\omega$ and rotating field with strength $H_r=54\,Oe$, which is perpendicular to stationary field.  Frequency of rotating field was fixed in range $0.05-2\,Hz$ and the constant field was changed in step like manner. In this experiment mixture of glycerol and water in volume fractions 3:7 were used.

Rods were tracked with image processing algorithm in Matlab. An ellipse was used as the descriptive shape of particles. The angle and length of major axis were chosen as the descriptive parameters and plotted along time axis. In cases when the ellipse was very short, the image was either considered faulty or the rod had flipped along z axis.

\begin{figure}
                \includegraphics[width=0.48\textwidth]{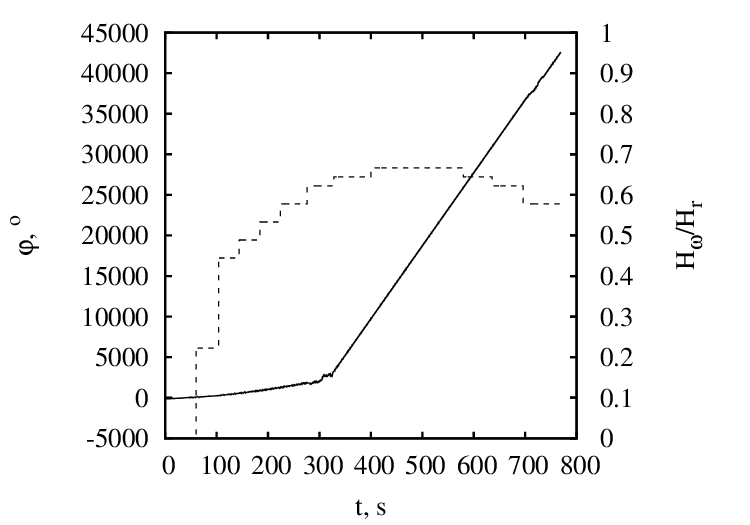}
                \caption{Angle of the particle $\varphi$ (left axis, solid line) and stationary field $H_\omega$ (right axis, dashed line) as function of time. Precessing field is constructed from stationary field $H_\omega$ and rotating field with strength $H_r=54\,Oe$ and frequency $f=0.25\,Hz$}
                \label{fig:JanusExp}
\end{figure}

Experimentally obtained angle of the rod as function of time, when stationary field is changed stepwise, is shown in Fig. \ref{fig:JanusExp}. It can be seen, that for small constant field strengths $H_\omega$ rod rotates slower than field frequency, because it is in asynchronous regime, but, when the stationary field strength $H_\omega$ increases above critical value, rod starts to rotate with field frequency.

\begin{figure}
                \includegraphics[width=0.48\textwidth]{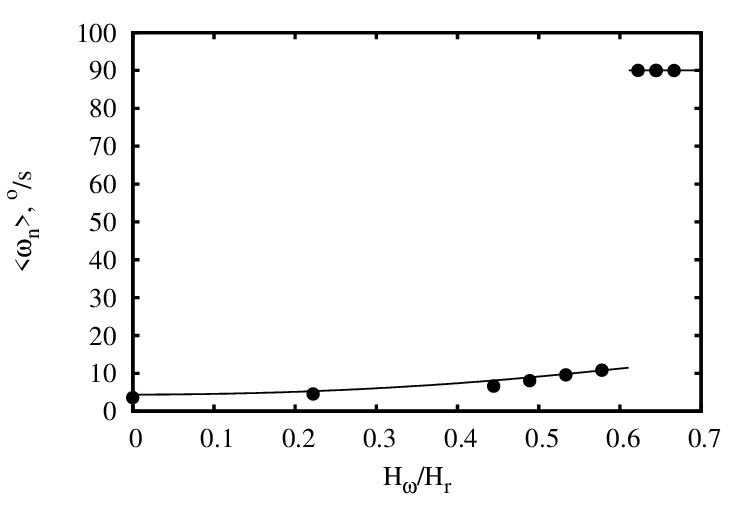}
                \caption{Mean rotational frequency $\langle\omega_n\rangle$ in dependence of constant field strength $H_\omega$. Rotating field strength is $H_r=54\,Oe$ and frequency is $f=0.25\,Hz$. Dots show experimentally obtained results and solid line is fit to theory.}
                \label{fig:omegan_exp}
\end{figure}

If we plot rotation frequency as function of stationary field strength $H_\omega$ we obtain Fig. \ref{fig:omegan_exp}. In Fig. \ref{fig:omegan_exp} is shown only angular velocities when particle stabilized its rotation. It is assumed that particle is stabilized its rotation after change of the field strength if fit to the straight line of $\varphi(t)$ slope has error less then 20\%. Otherwise some first points are omitted in fit. If fit has less then 10 periods and still is not stabilized, this field parameter is omitted in Fig. \ref{fig:omegan_exp}.

The value of $H_\omega$ where particle switch between synchronous and asynchronous regime we denote by critical field $H_c$. Knowing $H_c$, we can calculate that $\sigma^2=\frac{H_r^2}{H_r^2+H_c^2}$, when particle changes stability. Using neutral stability curve Eq. \eqref{eq:h13} or Eq. \eqref{eq:omega12}, which depends on critical $\sigma$ value, we can calculate anisotropy frequency $\omega_a$ of the particle. Anisotropy frequency depends on magnetic field strength, fluid viscosity and form of the particle.
We will use that hydrodynamic anisotropy axis coincide with magnetic anisotropy axis, therefore $\xi_r$ (for rotation about the equatorial semiaxes) is given in \cite{CebersElipse}
\begin{equation}\label{eq:xiElipse}
  \xi=8\pi\eta V \Gamma\text{ ,}
\end{equation}
where $\eta$ is dynamic viscosity of the particle surrounding fluid, $V$ is volume of the particle and $\Gamma$ is geometric factor which depends on spheroid semi-axis $a$ and $b$:
\begin{equation}\label{eq:Gamma}
  \Gamma=\dfrac{a^2+b^2}{a^2 N_\|+b^2 N_\perp}\text{ ,}
\end{equation}
where $N_\|$ and $N_\perp$ are demagnetization factors \cite{OsbornDemag}.
We define formfactor $F$ of the particle as \cite{SoftMatter}
\[
F=\frac{\chi_\|-\chi_\perp}{8\pi\Gamma}=\frac{\omega_a \eta}{H^2}=\frac{\omega_a \eta}{H_r^2+H_c^2}
\]
Formfactor of the particle defines magnetic and hydrodynamic properties of the particle and depends only on the particle.

Obtained $H_c$ and calculated $\omega_a$ gives rotational frequency of the particle in asynchronous regime which is in good agreement to theory \eqref{eq:omegan} as can bee seen in Fig. \ref{fig:omegan_exp}.
 
\section{Conclusions}

Analytic expressions of spheroidal paramagnetic particle stable synchronous and asynchronous rotation in precessing magnetic field is found. Results are found in non-inertial coordinate system rotating with magnetic field, which synchronous rotation change to stable point but asynchronous rotation to limit circle, which can be found analytically.
It is shown that numerically obtained bifurcation diagram \cite{superCimurs} have analytic form.
Obtained results are applicable for both prolate and oblate spheroids.
It is shown that experimentally obtained results for prolate particle are in good agreement with theoretical predictions. 


\begin{acknowledgments}
This work was supported by  PostDoc Latvia\\(Project no.  1.1.1.2/VIAA/1/16/060)
\end{acknowledgments}

\appendix

\section{Cubic equation for finding fixed points}\label{apx:cubic}

We start with equation:
\begin{equation}\label{eq:dndtJanus}
  \dfrac{d\vec n}{dt}=\omega_a\left(\vec n\cdot\vec h\right)\left[\vec h-\left(\vec n\cdot\vec h\right)\vec n\right]+\omega_H\vec e_H\times\vec n
\end{equation}
 The fixed points are defined as $\dot{\vec n}=0$ which gives:
\begin{equation}\label{eq:Sinh2}
  -\omega_H[\vec e_H\times\vec n]=\omega_a(\vec n\cdot\vec h)[\vec h-(\vec n\cdot\vec h)\vec n]
\end{equation}
Scalar multiplication of \eqref{eq:Sinh2} by $\vec e_H$, $\vec h$ and $[\vec e_H\times\vec n]$ gives consequently:
\begin{equation}
  \vec h\cdot\vec e_H =(\vec n\cdot\vec h)(\vec n\cdot\vec e_H)\label{eq:sinh_1}\\
\end{equation}
\begin{equation}
  -\vec e_H\cdot[\vec n\times\vec h] =\frac{\omega_a}{\omega_H}(\vec n\cdot\vec h)\left[1-(\vec n\cdot\vec h)^2\right]\label{eq:sinh_2}\\
\end{equation}
\begin{equation}
  1-(\vec n\cdot\vec e_H)^2 =-\frac{\omega_a}{\omega_H}(\vec n\cdot\vec h)(\vec e_H\cdot[\vec n\times\vec h])\text{ .}\label{eq:sinh_3}
\end{equation}
By eliminating $(\vec e_H\cdot[\vec n\times\vec h])$ and $\vec n\cdot\vec e_H$ from \eqref{eq:sinh_1}, \eqref{eq:sinh_2} and  \eqref{eq:sinh_3} we get bi-cubic equation for $\vec n\cdot\vec h$:
\begin{equation}\label{eq:bi-cubic}
  (\vec n\cdot\vec h)^6-(\vec n\cdot\vec h)^4+\frac{\omega_H^2}{\omega_a^2}(\vec n\cdot\vec h)^2-\frac{\omega_H^2}{\omega_a^2}(\vec h\cdot\vec e_H)^2=0
\end{equation}
By introducing parameters $y=(\vec n\cdot\vec h)^2$, $\omega=\dfrac{\omega_H}{\omega_a}$ and $\sigma=\vec h\cdot\vec e_H$, we get the cubic equation
\begin{equation}\label{eq:cubic=0}
  f(y)=y^3-y^2+\omega^2  y-\omega^2 \sigma^2=0\text{ .}
\end{equation}

\section{Perturbation around fixed points}\label{apx:pert}

In order to investigate stability of synchronous regime we introduce small perturbation $\vec \varepsilon$ from the synchronous state $\vec n$. Since $\vec n$ is unit vector and its length can not change $\vec \varepsilon$ should be perpendicular to $\vec n$ and can be written in form:
\begin{equation}\label{eq:varepsilon1}
  \vec\varepsilon=\varepsilon_1[\vec n\times\vec h]+\varepsilon_2[\vec e_H\times\vec n]\text{ ,}
\end{equation}
where $\varepsilon_1$ and $\varepsilon_2$ are small perturbations in two non-collinear directions.

Giving small perturbation $\vec\varepsilon$ to \eqref{eq:dndtJanus} gives equation for perturbation in the linear form:
\begin{equation}\label{eq:dvarepsilondt_1}
  \begin{aligned}
  \dfrac{d\vec \varepsilon}{dt}&=\omega_a(\vec\varepsilon\cdot\vec h)(\vec h-\vec n(\vec n\cdot\vec h))
    -\omega_a\vec\varepsilon(\vec n\cdot\vec h)^2-\\
    &-\omega_a\vec n(\vec \varepsilon\cdot\vec h)(\vec n\cdot\vec h)+\omega_H \vec e_H\times\vec\varepsilon
    \end{aligned}
\end{equation}

%
Before we can go further, some relations should be computed, which comes from \eqref{eq:Sinh2} and its consequences \eqref{eq:sinh_1}, \eqref{eq:sinh_2} un \eqref{eq:sinh_3}:
\[
\begin{aligned}
  (\vec\varepsilon\cdot\vec h)&=\varepsilon_2(\vec e_H\cdot[\vec n\times\vec h])\overset{\text{\eqref{eq:sinh_2}}}{=}\\
    \overset{\text{\eqref{eq:sinh_2}}}{=}&\varepsilon_2\dfrac{\omega_a}{\omega_H}(\vec n\cdot\vec h)(1-(\vec n\cdot\vec h)^2)
\\
  \omega_a(\vec h-\vec n(\vec n\cdot\vec h))&
    \overset{\text{\eqref{eq:Sinh2}}}{=}\dfrac{\omega_H}{(\vec n\cdot\vec h)}[\vec e_H\times\vec n]
\\
  \Big[\vec e_H\times[\vec n\times\vec h]\Big]&=\Big(\vec n(\vec e_H\cdot\vec h)-\vec h(\vec e_H\cdot\vec n)\Big)
  \overset{\text{\eqref{eq:sinh_1}}}{=}\\
  \overset{\text{\eqref{eq:sinh_1}}}{=}&\Big(\vec n(\vec e_H\cdot\vec n)(\vec n\cdot\vec h)-\vec h(\vec e_H\cdot\vec n)\Big)
  \overset{\text{\eqref{eq:Sinh2}}}{=}\\
  \overset{\text{\eqref{eq:Sinh2}}}{=}&-\dfrac{\omega_H}{\omega_a}\dfrac{(\vec e_H\cdot\vec n)}{\vec n\cdot\vec h}[\vec e_H\times\vec n]
\end{aligned}
\]
\[
\begin{aligned}
  &\omega_a\vec n(\vec \varepsilon\cdot\vec h)(\vec n\cdot\vec h)+\varepsilon_2\omega_H\Big[\vec e_H\times[\vec e_H\times\vec n]\big]
  \overset{\text{\eqref{eq:sinh_3}}}{=}\\&\overset{\text{\eqref{eq:sinh_3}}}{=}
    \vec n \varepsilon_2\omega_H(1-(\vec n\cdot\vec e_H)^2)+\varepsilon_2\omega_H(\vec e_H(\vec e_H\cdot\vec n)-\vec n)=\\
  &=\varepsilon_2\omega_H (\vec e_H\cdot\vec n)(\vec e_H-\vec n(\vec e_H\cdot\vec n))=\\
  &=\varepsilon_2\omega_H (\vec e_H\cdot\vec n) \Big[\vec n\times[\vec e_H\times\vec n]\Big]
  \overset{\text{\eqref{eq:Sinh2}}}{=}\\&\overset{\text{\eqref{eq:Sinh2}}}{=}
    \varepsilon_2\omega_a(\vec e_H\cdot\vec n)(\vec n\cdot\vec h)[\vec n\times\vec h]
\end{aligned}
\]
From eqs. \eqref{eq:dvarepsilondt_1} and \eqref{eq:varepsilon} using calculated relations we can separate terms with $[\vec n\times\vec h]$ and $[\vec e_H\times\vec n]$. We get equations for $\varepsilon_1$ and $\varepsilon_2$ time derivatives:
\begin{equation}\label{eq:e1e2}
  \left\{\begin{aligned}
  \dfrac{d\varepsilon_1}{dt}&=-\omega_a (\vec n\cdot\vec h)^2 \varepsilon_1
    -\omega_a(\vec e_H\cdot\vec n)(\vec n\cdot\vec h) \varepsilon_2\\
  \dfrac{d\varepsilon_2}{dt}&=\dfrac{\omega_H^2}{\omega_a}\dfrac{(\vec e_H\cdot\vec n)}{(\vec n\cdot\vec h)}\varepsilon_1
    +\omega_a\left(1-2(\vec n\cdot\vec h)^2\right)\varepsilon_2
  \end{aligned}\right.
\end{equation}

The stability of the fixed point is determined by eigenvalues of Jacobi matrix $\mathbf{J}$, where $\mathbf{J}$ is seen in \eqref{eq:e1e2}
\[
  \left(\begin{array}{c}{\dot \varepsilon_1}\\{\dot \varepsilon_2}\end{array}\right)=\mathbf{J} \left(\begin{array}{c}{ \varepsilon_1}\\{\varepsilon_2}\end{array}\right)
\]
The eigenvalues of Jacobi matrix are:
\begin{equation}\label{eq:lambda1}
  \lambda_{1,2}=\omega_a\dfrac{\tau\pm\sqrt{\tau^2-4\Delta}}{2}\text{ ,}
\end{equation}
where
\begin{equation}\label{eq:tau}
  \tau=1-3(\vec n\cdot\vec h)^2
\end{equation}
is trace of $J$ divided by $\omega_a$ and
\begin{equation}\label{eq:Delta}
  \Delta=\left(\frac{\omega_H}{\omega_a}\right)^2-2(\vec n\cdot\vec h)^2+3(\vec n\cdot\vec h)^4
\end{equation}
is determinant of $J$ divided by $\omega_a^2$, where $(\vec e_H\cdot\vec n)$ is excluded using \eqref{eq:sinh_2} and \eqref{eq:sinh_3}.

%
%

\section{Limit circle}\label{apx:circle}

In $\omega$ and $\sigma$ region where no fixed points can be found, there should be limit circle. We assume that without fixed points, $\vec n$ will rotate around some vector $\vec e_L$ with time varying frequency $\omega_L$. We assume that $\vec e_L\perp\vec n$. So we can write $\omega_L\vec e_L=\vec n\times\dot{\vec n}$. Here we introduce one more unit vector $\vec e_n$ with propriety $\dot{\vec n}=\left|\dot{\vec n}\right|\vec e_n$. This gives relation between 3 unit vectors $\vec e_L=\vec n \times\vec e_n$.

We need to find how $\vec e_L$ changes:
\begin{equation}\label{eq:basis1}\begin{aligned}
 \dfrac{d\vec e_L}{dt}&=\dfrac{d\left(\vec n\times e_n\right)}{dt}=\dfrac{d\vec n}{dt}\times \vec e_n+\vec n\times\dfrac{d\vec e_n}{dt}=\\
 &=\vec n\times\dfrac{d}{dt}\left(\dfrac{\dot{\vec n}}{\left|\dot{\vec n}\right|}\right)=
 \\
 &=\dfrac{1}{\left|\dot{\vec n}\right|}\vec n\times\dfrac{d^2\vec n}{dt^2}+\dfrac{d}{dt}\left(\dfrac{1}{\left|\dot{\vec n}\right|}\right)\left|\dot{\vec n}\right|\vec n\times \vec e_n\text{ ,}
 \\
 &=\dfrac{1}{\left|\dot{\vec n}\right|}\vec n\times\dfrac{d^2\vec n}{dt^2}-\dfrac{1}{\left|\dot{\vec n}\right|}\left(\vec e_n\cdot\dfrac{d^2\vec n}{dt^2}\right)\vec n\times \vec e_n\text{ .}
\end{aligned}\end{equation}
Since vectors $\vec n$, $\vec e_L$ and$\vec e_n$ are perpendicular we can express $\dfrac{d^2\vec n}{dt}$ is basis of these vectors as
\[
  \dfrac{d^2\vec n}{dt}=\left(\dfrac{d^2\vec n}{dt}\cdot \vec n\right)\vec n+\left(\dfrac{d^2\vec n}{dt}\cdot \vec e_L\right)\vec e_L+\left(\dfrac{d^2\vec n}{dt}\cdot \vec e_n\right)\vec e_n\text{ .}
\]
This simplifies \eqref{eq:basis1} to
\begin{equation}\label{eq:basis2}
  \dfrac{d\vec e_L}{dt}=-\dfrac{1}{\left|\dot{\vec n}\right|}\left(\dfrac{d^2\vec n}{dt^2}\cdot\vec e_L\right)\vec e_n\text{ .}
\end{equation}
Next we expand from \eqref{eq:dndtJanus}:
\[\begin{aligned}
  \dfrac{d^2\vec n}{dt}=&\omega_a\left(\dfrac{d\vec n}{dt}\cdot\vec h\right)\left[\vec h-2(\vec n\cdot\vec h)\vec n\right]-\\
  &-\omega_a(\vec n\cdot\vec h)^2\frac{d\vec n}{dt}+\omega_H\vec e_H\times\frac{d\vec n}{dt}
\end{aligned}\]
and get that
\[
  \dfrac{1}{\left|\dot{\vec n}\right|}\left(\dfrac{d^2\vec n}{dt^2}\cdot\vec e_L\right)=\omega_a(\vec e_n\cdot\vec h)(\vec h\cdot\vec e_L)+\omega_H\left[\vec e_H\times\vec e_n\right]\cdot\vec e_L\text{ .}
\]
Putting this in \eqref{eq:basis2} gives
\begin{equation}\label{eq:basis3}
  \dfrac{d\vec e_L}{dt}=-\omega_a(\vec e_n\cdot\vec h)(\vec h\cdot\vec e_L)\vec e_n-\omega_H\left(\vec e_H\cdot\vec n\right)\vec e_n\text{ .}
\end{equation}

Multiplying \eqref{eq:dndtJanus} by $\vec e_L$ and requiring that $\dfrac{d\vec n}{dt}$ is perpendicular to $\vec e_L$ gives that:
\begin{equation}\label{eq:dndteL}
  \omega_a(\vec n\cdot\vec h)(\vec h\cdot\vec e_L)=-\omega_H(\vec e_H\times\vec n)\cdot\vec e_L=\omega_H\vec e_H\cdot\vec e_n
\end{equation}
by adding and subtracting \eqref{eq:dndteL} multiplied by $\vec n$ to eq. \eqref{eq:basis3} we get:
\begin{equation}\label{eq:basish}\begin{aligned}
  \dfrac{d\vec e_L}{dt}=&-\omega_a(\vec e_n\cdot\vec h)(\vec h\cdot\vec e_L)\vec e_n-\omega_H\left(\vec e_H\cdot\vec n\right)\vec e_n+\\
  &+\omega_H\left(\vec e_H\cdot\vec e_n\right)\vec n-\omega_a(\vec n\cdot\vec h)(\vec h\cdot\vec e_L)\vec n=
  \\
  =&-\omega_a\left(\vec h\cdot\vec e_L\right)\left[\left(\vec e_n\cdot\vec h\right)\vec e_n+\left(\vec n\cdot\vec h\right)\vec n\right]+\\
  &+\omega_H\left[\left(\vec e_H\cdot\vec e_n\right)\vec n-\left(\vec e_H\cdot\vec n\right)\vec e_n\right]
\end{aligned}\end{equation}
We can divide $\vec h$ in the basis $\vec n$, $\vec e_n$, $\vec e_L$:
\[
  \vec h=\left(\vec n\cdot\vec h\right)\vec n+\left(\vec e_n\cdot\vec h\right)\vec e_n+\left(\vec e_L\cdot\vec h\right)\vec e_L
\]
and we get that
\[
  \vec h-\left(\vec e_L\cdot\vec h\right)\vec e_L=\left(\vec n\cdot\vec h\right)\vec n+\left(\vec e_n\cdot\vec h\right)\vec e_n
\]
Second term in \eqref{eq:basish} is double product:
\[
  \left(\vec e_H\cdot\vec e_n\right)\vec n-\left(\vec e_H\cdot\vec n\right)\vec e_n=\vec e_H\times[\vec n\times\vec e_n]=\vec e_H\times\vec e_L
\]
Putting all in \eqref{eq:basish} gives that:
\begin{equation}\label{eq:limCircle}
  \dfrac{d\vec e_L}{dt}=-\omega_a\left(\vec e_L\cdot\vec h\right)\left[\vec h-\vec e_L\left(\vec e_L\cdot\vec h\right)\right]+\omega_H\vec e_H\times\vec e_L
\end{equation}

This equation is similar to eq. \eqref{eq:dndtJanus} with opposite sign at $\omega_a$. This means that fixed points and their stability analysis for $\vec e_L$ is similar to the analysis done for $\vec n$.

In the region, where $\vec n$ does not have fixed point, it has limit circle and $\vec n$ rotates around circle orthogonal to $\vec e_L$.

\section{Rotational period}\label{apx:period}

We can write that $\vec n$ rotates:
\[\begin{aligned}
  \omega_L(t)\vec e_L&=\vec n\times\dfrac{d\vec n}{dt}=\\
  &=\omega_a\left(\vec n\cdot\vec h\right)\left[\vec n\times\vec h\right]+\omega_H\left[\vec e_H-\vec n\left(\vec e_H\cdot\vec n\right)\right]\text{ ,}
\end{aligned}\]
where $\omega_L(t)$ is angular velocity of $\vec n$ around $\vec e_L$.
Multiplying it by $\vec e_L$ gives that
\begin{equation}\label{eq:omegaL1}
  \omega_L(t)=\omega_a\left(\vec n\cdot\vec h\right)\left(\vec h\cdot\vec e_n\right)+\omega_H\left(\vec e_H\cdot\vec e_L\right)
\end{equation}

\begin{figure}[ht]
                \includegraphics[width=0.25\textwidth]{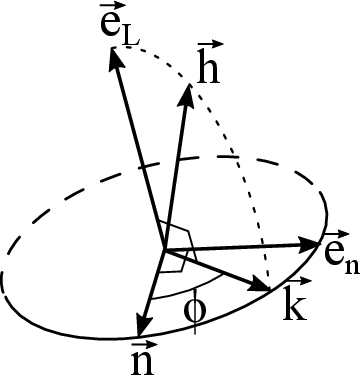}
                \caption{Vector $\vec k$ and angle $\phi$}
                \label{fig:veck}
\end{figure}

Next we introduce unit vector $\vec k$ which is in the direction of the projection of the vector $\vec h$ onto the plane, where $\vec n$ rotates and angle between $\vec k$ and $\vec n$ is $\phi$ (shown in fig. \ref{fig:veck}). So we can write that
\[
  \vec n\cdot\vec h=\left(\vec n\cdot\vec k\right)\left(\vec k\cdot\vec h\right)=\left(\vec k\cdot\vec h\right)\cos\phi
\]
\[
  \vec e_n\cdot\vec h=\left(\vec e_n\cdot\vec k\right)\left(\vec k\cdot\vec h\right)=\left(\vec k\cdot\vec h\right)\sin\phi
\]
\[
  \left(\vec k\cdot\vec h\right)^2=1-\left(\vec e_L\cdot\vec h\right)^2
\]
which simplifies eq. \eqref{eq:omegaL}
\begin{equation}\label{eq:omegaL(t)}
  \omega_L(t)=\dfrac{d\phi(t)}{dt}=\omega_a(\vec k\cdot\vec h)^2\sin\phi(t)\cos\phi(t)+\omega_H\left(\vec e_H\cdot\vec e_L\right)
\end{equation}
Separation of the variables and integration over the period gives:
\begin{equation}
  T=\dfrac{2\pi}{\sqrt{\omega_H^2(\vec e_L\cdot\vec e_H)^2-\frac{\omega_a^2}{4}\left(\vec k\cdot\vec h\right)^4}}
\end{equation}
or we can write that mean angular velocity is:
\begin{equation}
  \bar \omega_L=\sqrt{\omega_H^2(\vec e_L\cdot\vec e_H)^2-\frac{\omega_a^2}{4}\left(\vec k\cdot\vec h\right)^4}
\end{equation}
Since $\vec e_L$ fulfils the same dynamic equation as $\vec n$, then stable fixed points can be found from the same cubic eq. \eqref{eq:cubic}. Putting solutions of \eqref{eq:cubic} into mean velocity and using relations $y=(\vec e_L\cdot\vec h)^2$ and $(\vec e_L\cdot\vec h)(\vec e_L\cdot\vec e_H)=(\vec e_H\cdot\vec h)^2=\sigma^2$, we get:
\[
  \bar \omega_L=\sqrt{\omega_H^2\dfrac{\sigma^2}{y}-\frac{\omega_a^2}{4}(1-y)^2}=\omega_a \sqrt{\frac{\omega^2\sigma^2}{y}-\frac{1}{4}(1-y)^2}
\]
Using eq. \eqref{eq:cubic} it can be rewritten in form:
\begin{equation}\label{eq:omega_L_vid}
  \bar \omega_L=\omega_a \sqrt{\omega^2-\frac{1}{4}+\frac{y}{2}\left(\frac{3y}{2}-1\right)}\text{ ,}
\end{equation}
where $y$ is the real root of \eqref{eq:cubic}.

Solving \eqref{eq:omegaL(t)} we can also write how $\phi(t)$ changes:
\begin{widetext}
\begin{equation}
  \phi(t)=\arctan\left(\frac{2\bar\omega_L\tan\left[\omega_L(t-t_0)\right]-2\omega_H\left(\vec e_L\cdot\vec e_H\right)}{\omega_a\left(\vec k\cdot\vec h\right)^2}\right)+\pi k\text{ ,}
\end{equation}
\end{widetext}
where $t_0$ is some arbitrary constant and by changing $k\in{\rm I\!R}$ we can fulfil continuity of $\phi(t)$.


\bibliography{bib}

\end{document}